\documentstyle[12pt]{article}
\begin{document}
\baselineskip=20pt

\title{\LARGE Chiral Anomaly at Infinite Temperature}
\author{Henrique Boschi-Filho$^{\ast}$\\
\\
Instituto de F\'{\i}sica - Universidade Federal do Rio de Janeiro \\
Cidade Universit\'aria - Ilha do fund\~ao - Caixa Postal 68528 \\
21945-970 Rio de Janeiro, BRAZIL.}

\bigskip

\date{ }
\maketitle
\begin{abstract}
Using the heat kernel regularization we show that the Abelian chiral
anomaly in the limit of infinite temperature it is not a well defined
quantity, contrary to what happens at any finite temperature.  We show
that there is an ambiguity in the ordering of the limits of infinite
temperature and removal of the cut-off so that changing this ordering we
find different results for the chiral anomaly. We discuss these cases and
their possible interpretation.

\end{abstract}
\par

\bigskip
\bigskip


\vfill
\noindent $^\ast$e-mail: boschi@ufrj.bitnet\par
\hskip 1cm boschi@vms1.nce.ufrj.br\par
\pagebreak

\begin{section} {Introduction}

The chiral anomaly consists in the anomalous divergence of the
axial-vector current $\bar\psi\gamma_\mu\gamma_5\psi$, which is
classically conserved but not at the quantum level. It was originally
calculated by Schwinger \cite{Schwinger} and was later rediscovered
by Adler, Bell and Jackiw \cite{ABJ} studying the Veltman-Shutherland
paradox through the current algebra, in connection with the
$\pi^0\rightarrow 2\gamma$ decay \cite{Treiman}.

Since then, anomalies have played a special role in quantum field
theory because chiral anomalies mean violation of standard
Ward-Takahashi identities which are the guide for renormalizable
theories. The presence of chiral anomalies can be circumvented by
considering fermion families in which they cancel as a whole.
Particularly, this is the case of the Weinberg-Salam electroweak
theory. Another remarkable feature in this theory, is the fact that,
to warrant renormalizability, as is well known, the gauge fields are not
genuine massive but acquire mass through the spontaneous breakdown of the
vacuum symmetry.

In the beginning of seventies, Kirzinits and Linde \cite{Kirzinits}
proposed that the spontaneous symmetry breakdown mechanism may be
associated with the cooling of the universe and this could be
understood as a phase transition which had occurred at a certain
critical temperature.

The behavior of chiral anomalies at finite temperature was
first investigated by Dolan and Jackiw \cite{DJ} who showed that in
the case of massless two dimensional quantum electrodynamics the
anomalous divergence of the axial current was independent of any
finite temperature. Then, this result was rederived by other methods
and shown to be valid even in the non-Abelian case in four and
arbitrary (even) dimensions \cite{Mueller}.  Particularly, this was
done recently using the heat kernel regularization \cite{boschi1}
with the Fujikawa's method \cite{Fujikawa}, \cite{Reuter}. This
method was also applied successfully to models with curved background
fields \cite{boschi2}.

The relevance of the infinite temperature limit in quantum field theory is
manifold. First of all, this limit is considered in the study of the very
early universe as time approaches zero \cite{Linde}. Secondly, it is very
useful in dimensional reduction which is a requisite for compactification
on string theory \cite{Cremmer}. Finally, Polyakov has considered this
limit to show the existence of a deconfined phase in QCD \cite{Polyakov1}.
He had also shown that multinstanton configurations in $2+1$ dimensional
QCD could be associated with negative temperatures \cite{Polyakov2}, which
are usually recognized as temperatures ``higher'' than infinity in
paramagnetic systems.

In this paper we are going to discuss the infinite temperature limit for
the Abelian chiral anomaly. This is done using the heat kernel
regularization at finite temperature which we review in section 2.  In
section 3 we take the infinite temperature limit on the chiral anomaly, as
calculated in the preceeding section.  When we take the infinite
temperature limit we find that the ordering of this limit and the removal
of the cut-off is ambiguous, {\it i.  e.}, changing this ordering we are
lead to different results for the chiral anomaly. When the cut-off is
firstly removed we find the usual result for the chiral anomaly, but if we
first take the infinite temperature limit we find an ill defined quantity
which must be carefully treated. This is done with the help of a
$\zeta$-function prescription and alternatively with a $\theta$-function
transformation. A numerical approach is also sketched at the end of
section 3.  Discussion and conclusions are left for section 4.

\end{section}

\begin{section}{The anomaly at non-zero temperature}

Let us briefly review the paths leading to the Abelian chiral anomaly at
finite temperature $T$ which are essential to the discussion of its
infinite temperature limit.

In order to discuss non-zero temperature effects at equilibrium
conditions we start with a theory in Euclidean time restricted to the
interval $(0,\beta=1/T)$.  With this prescription and the
(anti)periodicity of the (fermi) boson fields we can map generating
functionals, in the absence of sources, on partition functions.

Fujikawa \cite{Fujikawa} showed that the chiral anomaly can be obtained
non-perturbatively in the path integral approach, observing that the
functional measures ${\cal D}\psi$ and ${\cal D}\bar\psi$ of the
fermi\-onic fields are not invariant under the infinitesimal chiral
transformations

$$\psi (x)\rightarrow \psi^\prime (x) = e^{i \alpha (x)\gamma_5} \psi
(x) $$

$$\bar\psi (x)\rightarrow \bar\psi^\prime (x) = \bar\psi (x) e^{i \alpha
(x)\gamma_5} $$

\noindent where $\gamma_5 = i \gamma_0\gamma_1\gamma_2\gamma_3$. This
noninvariance may be expressed through a Jacobian which can be
calculated expanding the fields $\psi\ (\bar\psi )$ and $\psi^\prime\
(\bar\psi^\prime )$ over complete orthogonal basis $\{\phi_n^\beta \}\
(\{{\phi_n^\beta}^\dagger\} )$, which are eigenstates of the Dirac operator
${{\slash\!\!\!\!D}} \phi_n^\beta =\lambda_n \phi_n^\beta$ and are antiperiodic
in the Euclidean time interval $(0,\beta)$. At finite temperature
$\beta^{-1}$, this Jacobian is given by
\cite{Reuter},\cite{boschi1}

$$ {\cal J}_\beta = \exp \{ -2i \int_0^\beta d\tau \int d^3 x\ \alpha
(x) {\cal A}_\beta (x)\}$$

\noindent where ${\cal A}_\beta (x)$ is an ill defined sum, usually called
anomaly,

$$ {\cal A}_\beta (x) = \sum_n {\phi_n^\beta}^\dagger (x) \gamma_5
\phi_n^\beta (x)$$

\noindent which, after regularization leads to anomalous
Ward-Takahashi identities.

We are going to use the heat kernel regularization which consists in
the introduction of a damping factor $\exp\{ -\epsilon\lambda_n^2\}$ in
the expression of the anomaly, to cut-off the high frequencies leading to
its regularized expression

\begin{eqnarray}
{\cal A}_\beta^R (x)
&=& \lim_{\epsilon\rightarrow 0} \sum_n {\phi_n^\beta}^\dagger (x) \gamma_5
\exp\{ -\epsilon\lambda_n^2\} \phi_n^\beta (x) \nonumber\\
&=&  \lim_{\epsilon\rightarrow 0} \sum_n {\phi_n^\beta}^\dagger (x) \gamma_5
\exp\{ -\epsilon\; {{\slash\!\!\!\!D}}^2\} {\phi_n^\beta} (x)\nonumber\\
&=&  \lim_{\epsilon\rightarrow 0} \lim_{x^\prime\rightarrow x} tr [ \gamma_5
\exp\{ -\epsilon\; {{\slash\!\!\!\!D}}^2\} \delta^\beta (x-x^\prime ) ]
\nonumber\\
&=&  \lim_{\epsilon\rightarrow 0} \lim_{x^\prime\rightarrow x} tr [ \gamma_5
H_\beta(x,x^\prime ;\epsilon ) ]
\end{eqnarray}

\noindent where $H_\beta (x,x^\prime;\epsilon)$ is the finite
temperature heat kernel

\begin{equation}\label{hk}
H_\beta (x,x^\prime ;\epsilon )=
\exp\{ -\epsilon\; {{\slash\!\!\!\!D}}^2\} \delta^\beta (x-x^\prime )
\end{equation}

\noindent which is related to the zero temperature one by
\cite{boschi1}\cite{boschi2}

\begin{equation}
H_\beta (x,x;\epsilon ) = H(x,x;\epsilon ) S (\beta^2/\epsilon )
\end{equation}

\noindent where $S (\beta^2/\epsilon )$ contains all the temperature
dependence of the heat kernel $H_\beta (x,x;\epsilon )$:

\begin{equation}\label{series}
S (\beta^2/\epsilon ) = 1 + 2
\sum_{n=1}^\infty (-1)^n \exp\{ {-n^2 \beta^2 \over 4\epsilon} \}
\end{equation}

In order to calculate explicitly the anomaly one can invoke the
Schwinger-\-DeWitt \cite{DeWitt} ansatz for its diagonal part

\begin{equation}
 H(x,x;\epsilon ) = (4\pi\epsilon )^{-d/2} \sum_{m=0}^\infty a_m
(x) \epsilon^m
\end{equation}

\noindent where $d$ is the space-time dimension and $a_m (x)$ are the well
known Seeley's coefficients \cite{DeWitt}. Owing to the properties of
gamma matrices it is possible to define a matrix $\gamma_{d+1}$ with the
same features of $\gamma_5$, {\it i. e.}, anticommuting with all other
$\gamma$s, for every even dimensional space-time. This way, one can show
that

\begin{eqnarray}\label{finite}
 {\cal A}^R_\beta &=& \lim_{\epsilon\rightarrow 0} tr [\gamma_{d+1}
H_\beta (x,x;\epsilon ) ] \nonumber \\
&=& \lim_{\epsilon\rightarrow 0} tr \bigl[\gamma_{d+1} H
(x,x;\epsilon ) S (\beta^2/\epsilon )  \bigr]\nonumber\\
&=& (4\pi)^{-d/2} tr [\gamma_{d+1}\ a_{d/2}(x)]
\lim_{\epsilon\rightarrow 0}\; S (\beta^2/\epsilon )
\end{eqnarray}

\noindent For any finite temperature, which is equivalent to have
$\beta\not= 0$, the above expression can easily be calculated since

\begin{equation}\label{lim1}
\lim_{\epsilon\rightarrow 0} S (\beta^2/\epsilon ) = 1\;.
\end{equation}

\noindent This assures the final temperature independence of the
chiral anomaly, leading to

\begin{equation}\label{anomaly}
 {\cal A}^R_{\beta\not= 0} (x)
= (4\pi)^{-d/2} tr [\gamma_{d+1}\ a_{d/2}(x)] \equiv  {\cal A}^R (x)
\end{equation}

\noindent which is the zero temperature anomaly in d (even) dimensions
\cite{DJ}-\cite{boschi1}. If d was odd we would have found a null result.

Another way of seeing this result is noting that the limit
$\epsilon\rightarrow 0$ on the function $S(\beta^2/\epsilon)$ imposed
on the regularized anomaly as the removal of the cut-off is
equivalent to the zero temperature $\beta\rightarrow\infty$ one, for which
$S(\beta^2/\epsilon)$ is again equal to unity.

\end{section}

\begin{section}{The infinite temperature limit}

Now, let us discuss the infinite temperature limit of the chiral anomaly.
As we are going to show, depending on the ordering of infinite temperature and
removal of cut-off limits on the calculation of the chiral anomaly we get
different results. Let us study
these cases separately.

\begin{subsection}{Case A}

Firstly we analyze the case of the ordering of limits where we first
remove the regulator ($\epsilon\rightarrow 0$) and then let the
temperature approach infinity ($\beta\rightarrow 0$). In this case we
have:

\begin{eqnarray}
{\cal A}^R_{\beta \rightarrow 0} (x)
&=& \lim_{\beta\rightarrow 0}\left\{
\lim_{\epsilon\rightarrow 0} tr [\gamma_{d+1}
H_\beta (x,x;\epsilon ) ]\right\} \nonumber \\
&=& \lim_{\beta\rightarrow 0}\left\{
\lim_{\epsilon\rightarrow 0} tr \bigl[\gamma_{d+1} H
(x,x;\epsilon ) S (\beta^2/\epsilon )\bigr] \right\} \nonumber\\
&=& \lim_{\beta\rightarrow 0}\left\{
 (4\pi)^{-d/2} tr [\gamma_{d+1}\ a_{d/2}(x)]
\lim_{\epsilon\rightarrow 0}\;S (\beta^2/\epsilon )\right\}
\end{eqnarray}

\noindent
Note that in the last line the limit ${\epsilon\rightarrow
0}$ was taken only on the heat kernel $H(x,x;\epsilon)$ and this limit on
the function $ S(\beta^2/\epsilon)$ is given by Eq.(\ref{lim1})
above, so

\begin{eqnarray}
{\cal A}^R_{\beta \rightarrow 0} (x)
&=& \;\lim_{\beta\rightarrow 0} (4\pi)^{-d/2} tr [\gamma_{d+1}\ a_{d/2}(x)]
\nonumber\\
&=& (4\pi)^{-d/2} tr [\gamma_{d+1}\ a_{d/2}(x)]
\nonumber\\
&=& {\cal A}^R (x)
\end{eqnarray}

\noindent which is the same result as that for the anomaly at zero
temperature, Eq.(\ref{anomaly}), since the removal of the cut-off
left the anomaly temperature independent.

\end{subsection}

\begin{subsection}{Case B}

Now, let us examine the case of the ordering of limits where we take first the
infinite temperature limit and then remove the cut-off:

\begin{eqnarray}\label{ARbeta}
{\cal A}^R_{\beta \rightarrow 0} (x)
&=& \lim_{\epsilon\rightarrow 0}\
\left\{ \lim_{\beta\rightarrow 0}\ tr [\gamma_{d+1} H_\beta
(x,x;\epsilon ) ]\right\} \nonumber\\
&=& \lim_{\epsilon\rightarrow 0}\ \left\{
(4\pi)^{-d/2} tr [\gamma_{d+1} H(x,x;\epsilon ) ]
\lim_{\beta\rightarrow 0}\ S (\beta^2 /\epsilon) \right\}
\end{eqnarray}

\noindent Note that, in this case, the series (\ref{series}) seems to
be not convergent, oscillating between $\pm 1$

\begin{equation}\label{pm1}
\lim_{\beta\rightarrow 0}
S(\beta^2/\epsilon)\ = \ 1 + 2 \sum_{n=1}^\infty (-1)^n
\ \rightarrow \pm\ 1
\end{equation}

\noindent (It may be interesting to note that this oscillating series
appears also in the quantization of superstrings involving an
infinite tower of ghosts \cite{Kallosh}). To remedy this situation we
are going to employ two approaches: ({\it i}) relate the series
$S(\beta^2/\epsilon)$ with a $\zeta$-function which has a well
defined limit for $\beta\rightarrow 0$; and ({\it ii}) use a
$\theta$-function transformation to evaluate it.

\begin{subsubsection}{$\zeta$-function approach}

To relate the series $S(\beta^2/\epsilon)$ in the limit
$\beta\rightarrow 0$ with the $\zeta$-function \cite{Hardy}-\cite{WW}

\begin{equation}\label{zeta}
 \zeta\ (s) = \sum_{n=1}^\infty\ n^{-s} \hskip 3cm (Re\ s > 1)
\end{equation}

\noindent we recall the relation

\begin{eqnarray}\label{relation}
\sum_{n=1}^\infty (-1)^{n} n^{-s} &=& - \sum_{n=1}^\infty n^{-s}
+ 2 \sum_{n=1}^\infty  {(2n)}^{-s} \nonumber \\
&=& -\; (1 - 2^{1-s}) \zeta\ (s)
\end{eqnarray}

\noindent so that comparing Eqs. (\ref{pm1}) and (\ref{relation}) we
choose as a prescription
\begin{equation}\label{lim2}
 \lim_{\beta\rightarrow 0} S(\beta^2/\epsilon) = \lim_{s\rightarrow 0}
\bigl[ 1 - 2(1 - 2^{1-s}) \zeta\ (s) \bigr]
\end{equation}

\noindent which is still ill defined since the $\zeta$-function
(\ref{zeta}) is convergent for $Re\ s > 1$ only. This problem has a
well known solution which is that of rewriting it through an analytic
continuation \cite{Hardy}-\cite{WW}

\begin{equation}
\zeta_R (s) = {1\over 2} + {1\over s-1} + 2\;\int_0^\infty
{(1+t^2)^{-s/2} \over e^{2\pi t} - 1} \sin \left( s\; \tan^{-1} t \right)
dt
\end{equation}

\noindent for which the limit $s\rightarrow 0$ is well defined

$$ \zeta_R \; (0) = - 1/2\;,$$
\noindent so that substituting this result in Eq. (\ref{lim2}) we
show that the temperature dependent sum $S(\beta^2/\epsilon)$
vanishes as $\beta$ approaches zero

\begin{equation}\label{lim3}
 \lim_{\beta\rightarrow 0} S(\beta^2/\epsilon) = 0\;.
\end{equation}

\noindent Here, the analytic continuation of the $\zeta$-function can be
thought as a complementary regularization needed to handle the undefined
sum $S(\beta^2/\epsilon)$ in the limit $\beta\rightarrow 0$ (or
$T\rightarrow\infty$).

Now, substituting these results in Eq.(\ref{ARbeta}), we can find the
regularized anomaly for this case

\begin{eqnarray}\label{ARzeta}
{\cal A}^{R\zeta}_{\beta \rightarrow 0} (x)
&=& \lim_{\epsilon\rightarrow 0}\ \left\{
(4\pi)^{-d/2} tr [\gamma_{d+1} H(x,x;\epsilon ) ]
\lim_{\beta\rightarrow 0}\ S (\beta^2 /\epsilon) \right\} \nonumber\\
&=& \lim_{\epsilon\rightarrow 0}\ \left\{
(4\pi)^{-d/2} tr [\gamma_{d+1} H(x,x;\epsilon ) ] \lim_{s\rightarrow 0}\
\bigl[ 1 - 2(1 - 2^{1-s}) \zeta_R \ (s) \bigr] \right\} \nonumber\\
&=& 0
\end{eqnarray}

\noindent since the limit $\epsilon\rightarrow 0$ which removes the
regulator was taken after the limit $\beta\rightarrow 0$ of infinite
temperature.

\end{subsubsection}

\begin{subsubsection}{$\theta$-function transformation approach}

We have shown above that the series $S(\beta^2/\epsilon)$ in the limit
$\beta\rightarrow 0$ can be regularized through the use of a
$\zeta$-function. Now we are going to show that to evaluate this sum
we can alternatively use the $\theta$-function transformation \cite{WW}

\begin{equation}\label{theta}
\sum_{n=1}^\infty e^{-\alpha n^2} = -{1\over 2} + {1\over
2}\sqrt{\pi\over\alpha} + \sum_{m=1}^\infty e^{-\pi^2 m^2/\alpha}
\end{equation}

\noindent The idea here is to rewrite $S(\beta^2/\epsilon)$,
Eq.(\ref{series}), with the help of the above equation.
Note that the limit $\beta\rightarrow 0$ in (\ref{series}) is
equivalent  to $\alpha\rightarrow 0$ in Eq. (\ref{theta}), which
diverges because of the presence of the term $1/2\sqrt{\pi /\alpha}$.
As was done for the $\zeta$-function we put

\begin{eqnarray}
\sum_{n=1}^\infty (-1)^n e^{-\alpha n^2}
&=& - \sum_{n=1}^\infty e^{-\alpha n^2}
+ 2 \sum_{n=1}^\infty e^{-\alpha {(2n)}^2} \nonumber\\
&=& -\left[ - {1\over 2} + {1\over
2}\sqrt{\pi\over\alpha} + \sum_{m=1}^\infty e^{-\pi^2 m^2/\alpha}\right]
\nonumber\\
& &\;\;\;+2 \left[ - {1\over 2} + {1\over
2}\sqrt{\pi\over 4\alpha} + \sum_{m=1}^\infty e^{-\pi^2 m^2/4\alpha}\right]
\nonumber\\
&=& - {1\over 2} -
 \sum_{m=1}^\infty e^{-\pi^2 m^2/\alpha}
+2  \sum_{m=1}^\infty e^{-\pi^2 m^2/4\alpha}\nonumber\\
&=&  - {1\over 2} -
 \sum_{m=1}^\infty e^{-\pi^2 m^2/\alpha}
\left(1 - 2 e^{3\pi^2 m^2/4\alpha}\right)
\end{eqnarray}

\noindent Observe that the divergent terms of Eq. (\ref{theta}) for
each $\theta$-function transformation mutually cancel in the above
expression, so that identifying $\alpha=\beta^2/4\epsilon$, we can
adopt as an alternative prescription for calculating the divergent
series (\ref{series})

\begin{eqnarray}\label{SR}
S^R(\beta^2/\epsilon)  &=&
1+ 2\sum_{n=1}^\infty (-1)^n e^{- \alpha n^2} \nonumber\\
&=& 1+ 2\left[ - {1\over 2} -
 \sum_{m=1}^\infty e^{- 4\pi^2 m^2 \epsilon/\beta^2}
\left(1 - 2 e^{3\pi^2 m^2\epsilon/\beta^2}\right)\right]\nonumber\\
&=& - 2  \sum_{m=1}^\infty e^{- 4\pi^2 m^2 \epsilon/\beta^2}
\left(1 - 2 e^{3\pi^2 m^2\epsilon/\beta^2}\right)
\end{eqnarray}

\noindent so that the anomaly at infinite temperature within this
prescription is

\begin{eqnarray}\label{ARtheta}
{\cal A}^{R\theta}_{\beta \rightarrow 0} (x)
&=& \lim_{\epsilon\rightarrow 0}\ \left\{
(4\pi)^{-d/2} tr [\gamma_{d+1} H(x,x;\epsilon ) ]
\lim_{\beta\rightarrow 0}\ S^R (\beta^2 /\epsilon) \right\}\;.
\end{eqnarray}

\noindent As the series $S^R(\beta^2/\epsilon)$ is well behaved
in the limit $\beta\rightarrow 0$

\begin{equation}\label{lim4}
 \lim_{\beta\rightarrow 0} S^R(\beta^2/\epsilon) = 0\;,
\end{equation}

\noindent we find that

\begin{eqnarray}\label{ARtheta2}
{\cal A}^{R\theta}_{\beta \rightarrow 0} (x)
&=& 0\;,
\end{eqnarray}

\noindent reproducing the result obtained with the $\zeta$-function
approach. Note that in the expression (\ref{SR}) for $S^R
(\beta^2/\epsilon)$ one can see the naive behavior of anomalies since it
has the form $0\times\infty$ as $\beta$ approaches zero. However the
vanishing term goes faster to zero than the divergent one to infinity.

\end{subsubsection}

\end{subsection}

\begin{subsection}{Case C}

We have shown above that changing the ordering of limits
$\beta\rightarrow 0$ and $\epsilon\rightarrow 0$ we obtain different
results for the chiral anomaly. So a natural question is that if one
can accommodate these results in a single fashion. This is not an easy
task since the approaches for cases A and B above were quite
different.

However, it is possible to show that if one takes these
limits simultaneously, this corresponds to establishing a relation
between these parameters as, for example, $\xi=\beta^2/\epsilon$. So,
depending on the path in which the limits $\beta\rightarrow 0$ and
$\epsilon\rightarrow 0$ are reached we have different values of $\xi$
and also for the anomaly.  So cases A and B correspond to
$\xi\rightarrow\infty$ and $\xi\rightarrow 0$, respectively. For any
intermediate value of $\xi$ the series $S(\beta^2/\epsilon)$, Eq.
(\ref{series}) can be treated numerically, since for any finite and
non-zero $\xi$ the sum

\begin{equation}
S_N (\xi) =
1+ 2\sum_{n=1}^N (-1)^n e^{- n^2 \xi /4}
\end{equation}

\noindent is ordinarily summable and approaches $S(\beta^2/\epsilon)$ as
$N\rightarrow\infty$. For example, if $\xi=0.5$ and $N=40$ we have

\begin{equation}
S_{N=40} (\xi=0.5) \simeq 2.68238 \times 10^{-8}
\end{equation}

\noindent which leads to an anomaly value intermediate between ${\cal A}^R$
and ${\cal A}^{R\zeta}_{\beta\rightarrow 0}$ (or ${\cal
A}^{R\theta}_{\beta\rightarrow 0}$). Some other values for $S_N (\xi)$ are
given in Table 1. Note that the series is rapidly
convergent for $\xi \geq 1$.

\begin{table} 
\caption{Numerical values for $S_N (\xi)$}
\label{table}
\begin{center}
\begin{tabular}{|c||l|l|}\hline
\hline
$\xi$ &  $\;\;\;S_{N=10}\;(\xi)\;\;\;$    &  $\;\;\;S_{N=40}\;(\xi)\;\;\;$
\\ \hline
\hline

0.001 & 0.97287133747 &  0.66361623101\\ \hline

0.01 &  0.75931044053 &  0.01648789112\\ \hline

0.1  &  0.06173907523  & $9.95794274078*10^{-19}$\\ \hline

1 &  $ 3.66139425893 *10^{-4} $   &   $ 3.66139425893 *10^{-4} $ \\ \hline

4 &    0.30062579744           &    0.30062579744        \\ \hline

10 &    0.83592080900         &    0.83592080900  \\ \hline

100 &    0.99999999997           &   0.99999999997   \\ \hline
\end{tabular}
\end{center}
\end{table} 

\end{subsection}
\end{section}

\begin{section}{Discussion and Conclusions}

We have shown above that the chiral anomaly at the infinite temperature
limit is not a well defined quantity. First, because of an ambiguity
in the ordering of limits of infinite temperature and removal of the
cut-off. Two possible orderings were examined as cases A and B of
section 3 and shown to lead to different results  for the chiral
anomaly. In particular, case A leads to the usual anomaly and in case
B, where we first take the infinite temperature limit, we have found
that the already regularized anomaly becomes again ill defined. To
handle this problem we have used $\zeta$- and $\theta$-functions
prescriptions which lead to the anomaly cancellation.

Now, we are going to do some speculation about the interpretation of
these results. Case A seems to represent a genuine theory at finite
temperature for which we take the infinite temperature limit.  Since
the chiral anomaly (after the removal of the cut-off) is temperature
independent the heating of the thermal bath in which the theory is
immersed has no effect on it.

Case B, in which the infinite temperature limit was taken before the
removal of the cut-off, led to the cancellation of the anomaly and must
corresponds to a different, if any, physical situation. This sequence of
limits suggests that this could be interpreted as a genuine theory at
infinite temperature which is not the limit of an usual theory at finite
temperature as that discussed in case A. This situation may had occurred,
for example, in the primordial singularity of the very early universe,
where the temperature was perhaps infinite. But the connection between the
singular case B and the usual one A is not clear at all. Some effort in
this direction has been made as was shown in case C of section 3, where we
had considered the two vanishing limits on $\epsilon$ and $\beta$ to be
taken simultaneously. The parameter $\xi=\beta^2/\epsilon$ approximately
describes the intermediate cases between A and B. However this was not a
satisfactory discussion since only a rough numerical approach was given.

These calculations were done using the heat kernel regularization (and a
complementary $\zeta$- or $\theta$-function prescription) but it could be
done as well with other schemes such as the Pauli-Villars one or
dimensional regularization.  It is worth mentioning that the cancellation
of the chiral anomaly occurs in perturbation theory at finite cut-off
\cite{Soto}.  This may be related to the present calculations if one
assumes the infinite temperature limit to be taken before the removal of
the cut-off.

Throughout this letter we have not mentioned background gravitational
fields which are relevant for cosmological considerations. In fact the
results derived above are still valid in the case where a background is
present if it is at least quasistatic (and may also have small
anisotropies) in order to insure thermodynamic quasiequilibrium
conditions, since the calculation of the chiral anomaly in this case is
analogous to the flat space-time one even at non-zero temperatures
\cite{boschi2}.

{}From the topological point of view, we know that the chiral anomaly is an
invariant, even at finite temperatures, so what happens in the infinite
temperature limit? One way of answering this question is to state that it
must be invariant if and only if the topology of the space-time is also
unchanging.  To see this let us remember that at finite temperatures the
topology of $(3+1)$ space-time is just ${\cal R}^3\times {\cal S}^1$ with
the one-sphere having a radius $\beta$. When we discussed case B we take
the limit $\beta\rightarrow 0$ (before the removal of the cut-off)
reducing the space-time dimension to $(3+0)$, which is equivalent to a
time compactification. So we find a null result for the chiral anomaly
since we have arrived at an odd dimensional space-time. However, in case A
we obtain the usual anomaly taking first $\epsilon\rightarrow 0$. Then we
let the temperature approach infinity.  This case corresponds to an
anomaly calculated in an even dimensional space-time (since the final
result was obtained before the dimensional reduction). We believe that his
picture can be better understood if cases A and B could be described in an
unified way. This is presently under investigation.

\bigskip
\bigskip
\noindent{\bf ACKNOWLEDGEMENTS.}
The author would like to acknowledge
C. Wotzasek for calling his attention to this problem and for useful
discussions. The author was also benefitted from conversations with
C.P. Natividade, M.S. Alves, C. Farina, N.  Braga and S. Rabello.
This work was partially supported by CNPq - Brazilian agency.

\end{section}

\vfill
\pagebreak

\end{document}